\documentclass[
aip,
rsi,
reprint,
notitlepage
]{revtex4-1}

\usepackage{amsmath}
\usepackage{graphicx}
\usepackage{hyperref}
\usepackage{float}
\begin{document}

\title{Multiple Parameter Dynamic Photoresponse Microscopy for data-intensive optoelectronic measurements of van der Waals heterostructures}
\author{Trevor B. Arp}
\affiliation{Quantum Materials Optoelectronics Laboratory, Department of Physics and Astronomy, University of California, Riverside, CA, 92521, USA}

\author{Nathaniel M. Gabor}
\email{nathaniel.gabor@ucr.edu}
\affiliation{Quantum Materials Optoelectronics Laboratory, Department of Physics and Astronomy, University of California, Riverside, CA, 92521, USA}
\affiliation{Department of Materials Science and Engineering, University of California, Riverside, CA, 92521, USA}
\affiliation{Canadian Institute for Advanced Research, CIFAR Azrieli Global Scholar, MaRS Centre West Tower, 661 University Avenue, Toronto, Ontario ON M5G 1M1, Canada}

\date{\today}

\begin{abstract}

Quantum devices made from van der Waals (vdW) heterostructures of two dimensional (2D) materials may herald a new frontier in designer materials that exhibit novel electronic properties and unusual electronic phases. However, due to the complexity of layered atomic structures and the physics that emerges, experimental realization of devices with tailored physical properties will require comprehensive measurements across a large domain of material and device parameters. Such multi-parameter measurements require new strategies that combine data-intensive techniques - often applied in astronomy and high energy physics - with the experimental tools of solid state physics and materials science. We discuss the challenges of comprehensive experimental science and present a technique, called Multi-Parameter Dynamic Photoresponse Microscopy (MPDPM), that utilizes ultrafast lasers, diffraction limited scanning beam optics, and hardware automation to characterize the photoresponse of 2D heterostructures in a time efficient manner. Using comprehensive methods on vdW heterostructures results in large and complicated data sets; in the case of MPDPM, we measure a large set of images requiring advanced image analysis to extract the underlying physics. We discuss how to approach such data sets in general, and in the specific case of a graphene - boron nitride - graphite heterostructure photocell.

\end{abstract}

\maketitle

\section{Introduction}

Since the discovery of graphene, nanotechnologists have developed rapidly evolving techniques to engineer novel quantum devices from atomically thin materials such as hexagonal boron nitride (hBN) and transition metal dichalcogenides (TMDs).\cite{Novoselov_Geim_Discovery_G_2004, Meyer_hBN_Synthesis_Properties_2008, Fuhrer_TMD_realization_2007} These materials can be stacked vertically into van der Waals (vdW) heterostructures that combine the electronic properties of the constituent materials in unusual ways.\cite{Geim_van_der_Waals_heterostructures_2013, Duan_Review_van_der_Waals_heterostructures_2016} Much recent research has focused on combining and engineering 2D materials to create designer properties that result from length scale engineering - tuning the electronic properties by structuring the critical device length scales at or below the electron wavelength.\cite{Gabor_Metamaterias_2018, Xu_Excitons_TMD_heterobilayers_2018, Fang_interlayer_coupling_vdW_2014, Zhang_moire_superlattices_heterobilayers_2017, Song_Topological_Bands_Graphene_Superlattice_2015} In loose analogy to optical metamaterials, engineering sub-wavelength structure in these quantum metamaterials may give unprecedented access to quantum material properties, allowing us to engineer custom unit cells, topological bands and altered excited states. Intriguingly, length scale engineering of these materials may also allow us to tune interactions between charge carriers in the materials, creating novel correlated electronic phases.\cite{Ye_Superconductivity_2012, Cao_Pablo_Superconductivity_2018, Huang_Layer_depend_Ferro_2017, Jiang_2D_vdW_Magnets_2018, Liu_Exciton_Condensate_2017, Li_Exciton_superfluid_2017}

The proliferation of available 2D materials, the means to assemble high quality heterostructures, and theoretical proposals of emergent phenomena have led to a remarkable growth in the complexity of vdW heterostructure stacks.  From these innovations, diverse research avenues have been initiated, yet many challenges lie ahead. The new materials and metamaterials are increasingly complex, and understanding their behavior involves probing large multi-variable parameter spaces. Individual electronic transport or optical probes of solid-state physics may not be sufficient for comprehensive understanding of emergent complex behavior. In this work, we identify the challenges involved in measuring complex quantum materials (Section I), present a combined optical and electronic transport technique, MPDPM, to overcome these challenges (Section II), and discuss how to analyze MPDPM data and draw conclusions from an example MPDPM measurement (Section III).

\subsection{Challenge: Complex Behavior Involving Multiple Parameters}

\begin{figure}[t]
\centering
\includegraphics{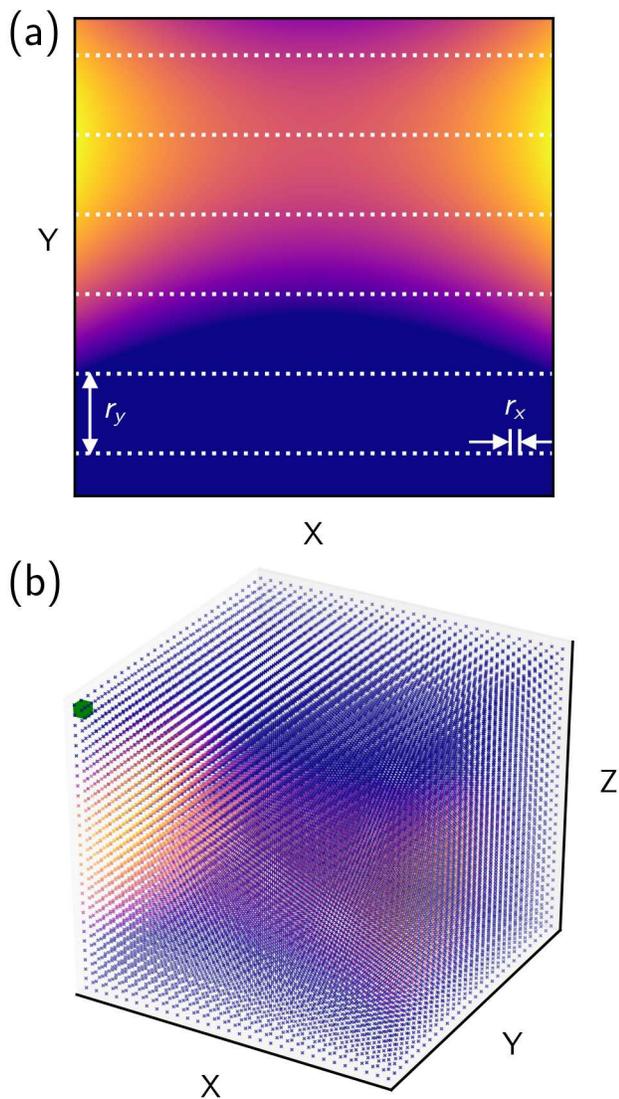}
\caption{
(a) Represents the phase space of a hypothetical phenomena that depends on two independent variables, with the observable value represented by a color scale. Single variable measurements are represented as dashed white lines with Y held constant. (b) Represents the phase space of a hypothetical phenomena that depends on three independent variables, each point in three-dimensional space has an observable value represented by color. The green cube in the upper left represents a single voxel.
}
\label{fig:phase_space}
\end{figure}

As nanotechnologists and materials scientists, how do we systematically assess complex electronic behavior that may arise in new material systems, particularly those with unusual synthetic properties? In solid-state physics, the answer has traditionally been to set up a single-parameter experiment that aims to cut through the complexity and capture quantum phenomena in as concise a measurement as possible. Typical experiments consist of well-established transport or spectroscopic measurements sampling over a \textit{single} independent variable. Often, these measurements use commercially available instruments. Implicit in this approach is the assumption that all other experimental parameters have negligible effect on the variable of interest. In 2D materials, many properties are the result of atomic thinness, which also makes them sensitive to external conditions, defying the assumption that other independent variables do not contribute to the electronic behavior. Truly comprehensive characterization using standard measurement approaches would require prohibitively long times, due in part to the measurement rate and the numerous trials required to address variations across many material parameters. As the complexity of 2D systems increases, new data intensive approaches - taking inspiration from astrophysics, high-energy physics, and biomedical imaging - must be developed.

In this section, we lay out an elementary assessment of the most restrictive experimental parameter - experimental time - and discuss how multi-variable searches can be optimized to improve the search for correlations across experimental variables. Fundamentally, experimental time T is the dominant limiting factor in measuring complex device behavior. Simply stated, the total time of a measurement combines the hardware-limited time per point $t_h$ with the sample response time $t_s$, multiplied by the total number of data points to be measured. 

To illustrate how the total time can be evaluated for a simple experimental system, Figure \ref{fig:phase_space}(a) shows a generic phenomenological response that depends on two experimental parameter dimensions, measured with single variable measurements. The experiment sweeps the X variable at constant Y, taking a series of line cuts through the experimental phase space. The time of such an experiment is given by $T = (t_h + t_s) \frac{\Delta X}{r_x} \frac{\Delta Y}{r_y}$, where $\Delta X$ and $\Delta Y$ are the ranges of X and Y defining the parameter space, and $r_x$ and $r_y$ are the resolutions of the X and Y variables. Generalizing to an N dimensional parameter space spanned by N independent variables, ($\mathbf{e}_{1}$, ..., $\mathbf{e}_{N}$): 
\begin{equation} 
\label{eqn:T}
T = (t_h + t_s) \prod_{i = 1}^{N} \frac{\Delta \mathbf{e}_{i}}{r_i}
\end{equation} 
Here, Equation \ref{eqn:T} can be understood intuitively as the time spent per voxel multiplied by the volume of the parameter space, $\prod_{i} \Delta \mathbf{e}_{i}$, divided by the voxel volume, $\prod_{i} r_i$. For a fixed parameter space volume, as the voxel volume decreases (i.e. the resolution increases), the total experimental time will increase.

Increasing the dimension N of a parameter space enforces greater limitations on total experimental time. To see this, Figure \ref{fig:phase_space}(b) illustrates the same phenomena as Figure \ref{fig:phase_space}(a) but in a 3D parameter space ($N = 3$), representing observables as colored points, and showing a voxel as a small green cube. Measurements of complex systems - those where non-trivial correlations exist between $N > 1$ independent variables - require significant values of $\Delta e$ and $r$ to obtain sufficient data for meaningful statistical analysis. In Figure \ref{fig:phase_space}(b), we see that due to the dimensionality of the phase space, the number of voxels is exponentially larger than for a two-dimensional experiment. Comprehensive measurements in a N-dimensional parameter space thus require exponentially more time.

High dimensional experimental phase spaces require making careful choices to minimize T while acquiring sufficient data for robust statistical analysis. Assume that, in general, T is large and constant, limited by experimenter (i.e., graduate student) time, sample lifetime or other resources. Optimizing high dimensional measurements involves optimizing the hardware, which decreases $t_h$, or optimizing the search of parameter space by making tradeoffs in $\Delta \mathbf{e}_{i}$ and $r_i$. However, the intrinsic sample response time $t_s$ limits how fast a measurement can proceed, and if $t_s >> t_h$, hardware optimization does little to increase measurement efficiency. Hardware optimization is application specific, we discuss it for heterostructures of 2D materials in Section \ref{sec:optics}. 

The greatest gains in efficiency come from tradeoffs in resolution. Ideally, the experimenter can reduce excessive resolution in one parameter to gain resolution in another parameter. Less ideally, the experimenter can choose to restrict the range of one or more parameter(s) $\Delta \mathbf{e}_{i}$, or neglect certain parameters, resulting in a narrower but better resolved measurement. The latter is the conventional strategy, which has greater likelihood of missing or misrepresenting phenomena occurring within a complex parameter space. 

In the large T limit, conventional single variable measurements are fundamentally inefficient. By their nature, single variable measurements explore one parameter, for example the X variable in Figure \ref{fig:phase_space}(a), with high resolution, and all other variables held constant, meaning $r_x << r_y, r_z, ..., r_N$. With hardware heavily optimized for only one variable it is difficult to effectively trade resolution in X for resolution in another variable and experimenters often deal with finite time by restricting the domains or omitting parameters. Single variable measurements become increasingly ineffective in identifying cross correlations between multiple parameters as the complexity of a measurement increases (i.e., as N increases), as higher resolution is needed, or as the relevant ranges become larger.

Does a better understanding of multi-parameter measurement science translate into accelerated discovery? While it is beyond the scope of this work, we posit that experimentalists using only standard techniques risk falling prey to a version of the availability heuristic. By focusing on measurements that are easy to perform with off-the-shelf or commercial equipment, complex phenomena that correlate across multiple parameters are missed or misinterpreted. Expectation bias is a danger when choosing parameters for new materials: an experimenter may unconsciously select the parameters that are most likely to conform to expectations or established models.\cite{Jeng_history_expectation_bias_physics_2005, Franklin_Selectivity_Discord_2002, Franklin_Bandwagon_1984, Mulargia_selection_bias_2001} Choosing which variables to hold constant can easily introduce selection bias that leads to compelling, yet incomplete, phenomenological knowledge complicating realistic interpretation. Comprehensive methods are therefore significantly advantageous in the search for new phenomena, particularly when a unique target system is probed using multiple non-standard experimental techniques.

\subsection{Challenge: Optoelectronic Measurements of 2D Materials}

\begin{figure}[t]
\centering
\includegraphics{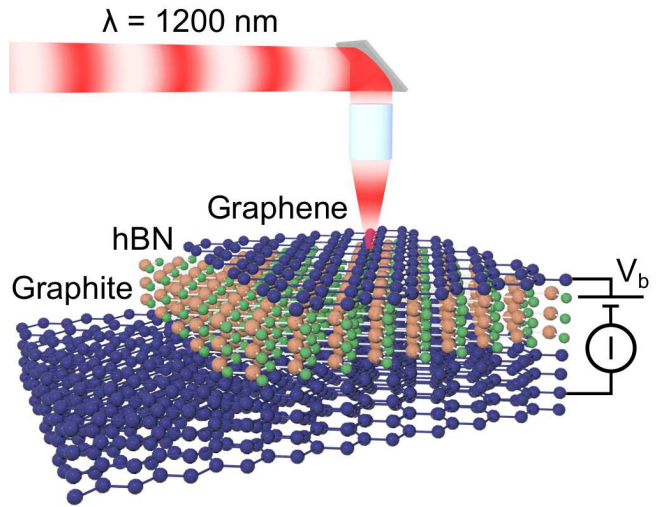}
\caption{Schematic of a vdW heterostructure made of graphene on top of hexagonal Boron Nitride (hBN) on top of ultra-thin graphite. The heterostructure is excited with a 1200 nm ultrafast pulsed laser and interlayer photocurrent, $I$, can be measured as a function of laser parameters (beamspot position, laser power, etc.) and bias voltage applied to the graphene, $V_{b}$.}
\label{fig:device}
\end{figure}

In optoelectronic materials, photo-excited electrons are promoted to high energies, leaving behind short-lived charge vacancies, or holes.  In this way, electrons promoted across a semiconductor band gap result in long-lived electron-hole pairs, while those excited in a semimetal may result in short-lived excitations. The timescale over which the electron-hole pairs recover to equilibrium is determined by energy and momentum relaxation processes in the material, which in turn depend on electronic band structure, electronic interaction strength, and electron-phonon coupling. 

In 2D semiconductors and semimetals, photoexcited electron-hole pairs may interact in unusual ways, giving rise to many body correlations that persist even at room temperature. In TMDs, charge carriers form hydrogen-like bound states with well-defined orbital and spin angular momentum.\cite{Mueller_Exciton_Review_2018, Wang_Excitons_review_2018} Depending on the structure of the material, these strongly bound excitons may be influenced by non-trivial bands such as topological or moir\'e bands, or have additional quantum numbers such as valley index or pseudospin.\cite{Gabor_Metamaterias_2018} In graphene, the electron-hole pairs form a rapidly evolving hot carrier distribution exhibiting unusual cooling pathways, with electron-electron and electron-phonon scattering processes competing to relax excess energy. Combining 2D semiconductors, 2D insulators, or semimetals into vdW heterostructures (such as the example graphene - boron nitride heterostructure, shown schematically in Figure \ref{fig:device}) introduces additional degrees of freedom, for instance allowing excitons to form with the electron and hole in different materials.\cite{Xu_Excitons_TMD_heterobilayers_2018} All of these unique properties contribute to energy and momentum relaxation, giving rise to highly complex behavior over a large range of time scales (from femtosecond electron-electron scattering to nanosecond exciton recombination).

These unusual electron-hole interactions in vdW metamaterials result in part from reduced dimensionality, which increases the energy scales of electronic states and interactions (e.g., increasing the binding energy of excitons).\cite{Chernikov_Exciton_WS2_2014} Due to electron confinement, 2D materials allow correlated or interacting phases to exist at higher temperatures than in conventional materials. Such effects are less accessible in 3D materials, which exhibit high symmetry due to translation invariance of the unit cell in all three spatial dimensions. Not only does high symmetry constrain the possible phenomena in many ways, it also allows the experimenter to make several assumptions about the behavior based on the unit cell. 2D materials inherently break several exploitable symmetries, expanding the space of possible phenomena and increasing the phase space for electronic states and interactions. 

In multiple respects the properties that make vdW heterostructure metamaterials interesting also make them difficult to measure and understand. Understanding electron-hole pair dynamics in 2D systems presents numerous experimental challenges since observable quantities - such as current, voltage, reflectivity or photoluminescence - are averaged in space and time. Purely electronic measurements only access low energy dynamics near the Fermi surface and average the electron dynamics over the spatial extent of the device. In the time domain, dynamics occur on timescales of femtoseconds to hundreds of picoseconds, and if an excitation persists significantly longer than those timescales, it will give only steady state equilibrium values. Gaining experimental information about the dynamics and testing theoretical models requires optical techniques with high spatial and/or temporal resolution.\cite{Heinz_Dynamics_heterostructures_2018} Moreover, in vdW heterostructures, multiple unusual electronic effects may overlap. Though individual effects could be exploited for manipulating electronic behavior, experiments must take into account and carefully control for all overlapping effects. Separating out individual properties requires multiple experimental variables, so that the property of interest can be uniquely accessed. 

\section{Multi-Parameter Dynamic Photoresponse Microscopy}

We describe a technique, called Multi-Parameter Dynamic Photoresponse Microscopy (MPDPM), that efficiently measures the optoelectronic response of vdW heterostructures. Utilizing diffraction limited optics, ultrafast lasers and scanning mirror optics, MPDPM excites the sample with a high intensity optical probe that drives the sample away from equilibrium, thus accessing correlated states, resolving short timescales, and producing high signal-to-noise photoresponse. The optical components are automated and controlled by an integrated, fully automated Data Acquisition (DAQ) program that simultaneously controls all other experimental parameters (such as applied voltage, magnetic field, temperature etc.). Such centralized control allows for efficient tradeoffs between parameters when exploring a large sample phase space. This technique acquires data rapidly, densely and systematically with respect to many experimental variables, resulting in high dimensional data arrays. The end result of MPDPM is a large set of photoresponse images spanning all relevant experimental variables that ideally capture all the complexity of device's phase space. Although the data sets are more complex than in conventional measurements, these large and complex data sets can be efficiently handled through careful data analysis, described in Section III.

\subsection{Diffraction Limited Ultrafast Optics}
\label{sec:optics}

\begin{figure}[t]
\centering
\includegraphics{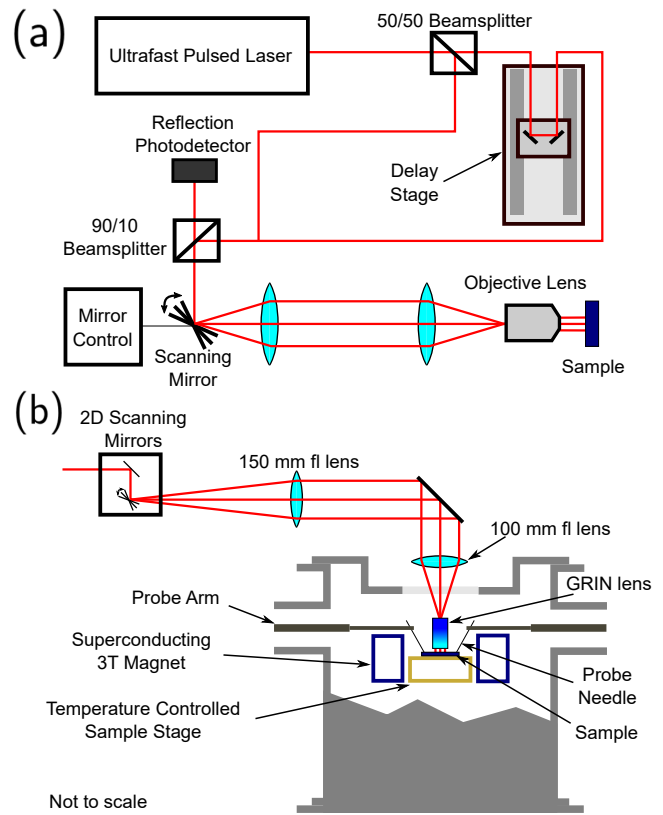}
\caption{Schematic of the optical setup (a) A diagram of the optics showing all the major optical components. (b) A cross sectional diagram of the optical setup and optical cryostat detailing the optics coupling into the GRIN lens.}
\label{fig:optics}
\end{figure}

\begin{figure}[t]
\centering
\includegraphics{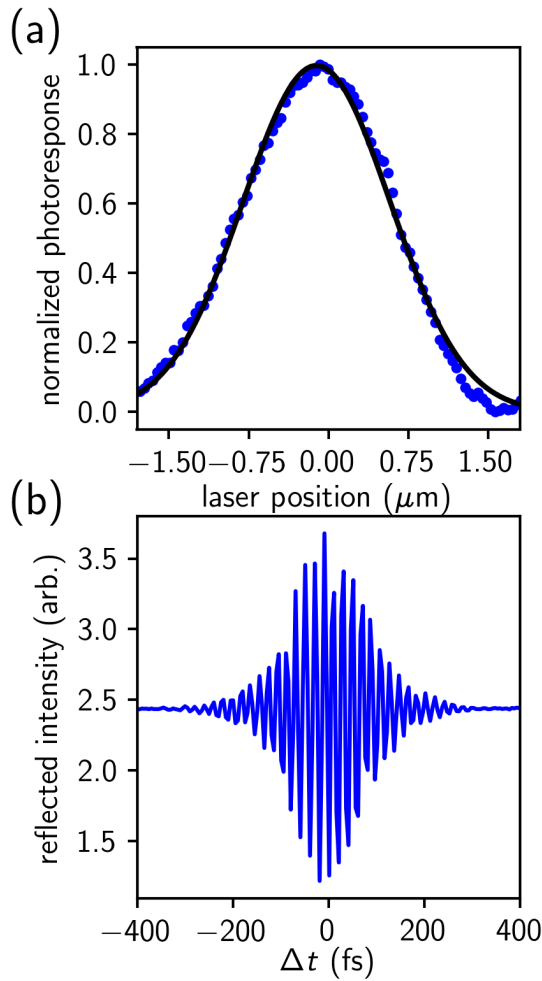}
\caption{Characterization of the ultrafast pulsed beam in the optical cryostat under vacuum. (A) Measured photoresponse of an absorber smaller than the diffraction limit. Black line is a fit to a Gaussian function with a full width at half max of 1.67~$\mu$m. (B) Two pulse autocorrelation as a function of the delay between two subsequent pulses, $\Delta t$.}
\label{fig:pulseoptics}
\end{figure}

MPDPM uses a local ultrafast optical probe to perform space-time resolved photocurrent and reflectance measurements. Incident light focused to the diffraction limit can resolve micron sized in-plane features, and the high incident intensity under a diffraction limited beamspot increases the signal and can drive the system well out of equilibrium. Using a scanning diffraction limited beamspot also allows light reflected back through the optics to be focused onto a single pixel detector, with much higher signal to noise than a CCD. The dynamics of charge carries often occur on timescales of order femtoseconds to picoseconds, so excitation by a continuous wave laser gives only equilibrium, steady state values, washing out the dynamics. Therefore, the optical probe must be localized in time as well as space. Ultrafast pulsed lasers can generate pulses on the order of the dynamics, giving access to phenomena that occur on those relevant timescales. In addition, the high peak pulse intensity increases the fluence of incident light, driving the system harder and increasing the signal.

To generate an optical probe that is local in space and time, we combine the techniques of scanning beam photocurrent and reflectance microscopy with ultrafast optical two-pulse measurements. \cite{Steinmeyer_Ultrafast_optics_Review_2003, Mizuho_application_pump_probe_2008} A schematic of the optical system is shown in Figure \ref{fig:optics}(a). We use a MIRA 900 OPO ultrafast laser which generates 150 fs pulses with controllable wavelength from 1150~nm to 1550~nm at a 76~MHz repetition rate. The output of the laser is split into two paths by a 50/50 beamsplitter and a translation stage is used to controllably introduce a path length difference. The two beams are then recombined, and the path length difference splits a single laser pulse into two sub-pulses separated by a time delay, $\Delta t$.

The recombined beam is fed into scanning beam optics which consist of rotating mirrors and a system of two lenses that focus the beam onto the back of an objective lens. The objective lens is set at the focal length of the second lens such that, as the scanning mirror rotates the beam is still focused onto the same position on the back of the objective, but arriving at different angles. The objective lens focuses the light down into a diffraction limited beamspot where the position of the beamspot depends on the incident angle. As the scanning mirror rotates, the beamspot moves over a large area of the sample without aberration, allowing for quick high-resolution scanning. Many conventional optoelectronic measurements keep the optics fixed and translate the sample. While simple, this technique is too slow to sample phase space in a time efficient way. When focused, the laser beamspot spatial profile is an Airy disk, which can be approximated using a Gaussian point spread function. Figure \ref{fig:pulseoptics}(a) shows the measured photoresponse of an absorber smaller than 1~$\mu$m using a wavelength of 1200~nm. The data is fit well by a Gaussian function (black line) with full width at half maximum of 1.67~$\mu$m, indicating that our system is at the diffraction limit.

Figure \ref{fig:optics}(b) details our specific scanning optics and the customized Janis Research ST-3T-2 optical cryostat that we use in our experiments. The sample sits in vacuum on a sample stage, which can controllably vary the temperature from 4~K to 420~K. The sample stage is in the center of a 3~Tesla superconducting magnet. The sample is electronically probed using four probe needles which contact conductive pads on quartz chip carriers that are wire-bonded to fabricated titanium-gold contacts on the sample. We then amplify the electrical signal and measure the current resulting from the incident laser light, or photocurrent. We also measure the reflectance of the sample by measuring the intensity of the light that is reflected from the sample with a near-infrared photodiode.

To fully enclose our focusing optics inside the vacuum chamber, we use a Gradient Index of Refraction (GRIN) lens as an objective. A GRIN lens is a single small cylinder of glass with the index of refraction varied radially. Lacking the many interfaces of a conventional objective, a GRIN lens does not disperse laser pulses as dramatically as a traditional objective. Figure \ref{fig:pulseoptics}(b) shows the autocorrelation of the reflected intensity due to two overlapping laser pulses, near $\Delta t = 0$. The autocorrelation width is approximately three times the pulse width. Our autocorrelation pattern is 570 fs wide, indicating that our pulses are 190 fs long at the sample, only 27\% off the 150~fs laser specification. Low dispersion allows us to measure short timescales and gives high peak pulse intensity. However, a GRIN lens also has downsides compared to a traditional objective lens. When well aligned, the power throughput of the GRIN lens is very high, however the process of aligning the optic over the sample under vacuum introduces systematic uncertainty into the laser power. Also, the field of view for a GRIN lens is typically smaller than a traditional objective lens, which is no problem for micron sized samples, but can limit applications in some large area samples. MPDPM can be performed using a traditional objective lens at the cost of increased pulse dispersion and therefore decreased time resolution and peak pulse intensity.

\subsection{Integrated Data Acquisition System}

\begin{figure*}[t]
\centering
\includegraphics{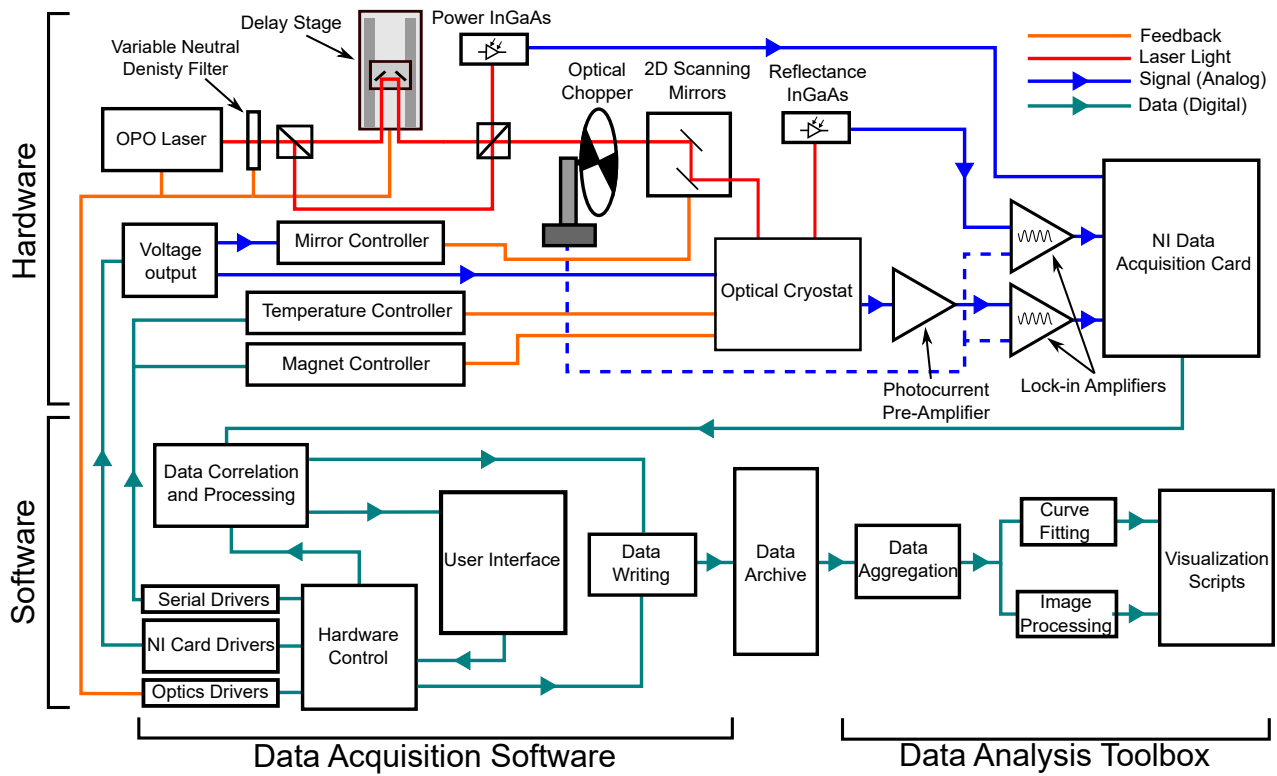}
\caption{Experimental data flowchart, schematically showing the flow of data between hardware and software components as well as the feedback involved in controlling the experiment.}
\label{fig:dataflow}
\end{figure*}

The goal of MPDPM is to time-efficiently sample as large of a parameter space as possible, using as many experimental parameters as are relevant and practical. To do this efficiently requires the ability to optimize the measurement time, as described in equation \ref{eqn:T}. The optics described in Section IIA are designed to allow fast scanning and other hardware components to be optimized to work as rapidly as possible, decreasing $t_h$ to a lower bound given by maximum hardware speed and amplifier time constants. Furthermore, the high signal-to-noise ratios can minimize $t_s$ to its intrinsic limit. Well-designed optics improve the time efficiency of the experiment ``for free." However, the largest increases in efficiency come from the ability to made tradeoffs in resolution. Optimal utilization of the optics and effective tradeoffs requires an integrated Data Acquisition (DAQ) system that automates all hardware components through one program. Such an integrated DAQ can control all hardware components at their optimum, in parallel, with minimal human input. The software allows the experimenter to choose the ranges and resolutions of various parameters in a scan in an intelligent manner, making appropriate tradeoffs. Finally, such a DAQ system allows data to be gathered densely, systematically, and repeatably, in a format that allows advanced data analysis.

We developed an integrated DAQ program using a set of python modules that interface with equipment drivers and control all hardware components simultaneously with the maximum amount of automation possible. Our experimental setup can scan a beam in two dimensions, while applying voltages to the sample under various optical conditions. In addition, the optical cryostat that contains our samples can control the temperature of the sample and apply a magnetic field. Each of these components requires specialized hardware, which were designed and selected to allow for full automation.  The flow of data is shown schematically in Figure \ref{fig:dataflow}. The main hardware components of the optics and controllers, shown in the upper left, are controlled with feedback to the DAQ software, which is represented in the lower left. From the user interface, any of the hardware components can be changed or scanned, varying some output over a given range. If one or two of the components is set to scan, the rest will be held constant. 

From the user interface the experimenter can define which parameters form the axes of a two dimensional scan, and define the scan's resolution in those parameters. The result is an array of data, or ``data plane." The experimenter can select a third parameter to scan over and the software will take successive 2D scans as a function of that parameter, constructing a 3D ``data cube" out of many data planes stacked along the third axis. These data planes or data cubes form a ``run," the discrete unit of MPDPM image data. In addition to the data, each run saves all possible control parameters, 125 in total, of the hardware and software to ensure consistency and repeatability. Each run is assigned a unique run number and the files for that run are saved to disk in a data archive. To efficiently take many runs, the software allows the user to repeat a run varying another parameter, taking data cubes as a function of this fourth parameter. Put together, the runs form a four dimensional ``data hypercube," sampling a large volume of parameter space. This allows the experimenter to, with full control over the ranges and resolutions of all parameters, efficiently sample a four-dimensional parameter space fully automatically. These fully automatic measurements can run for hours or days, collecting data with no human input needed.

\subsection{Data Taking for 2D Materials}

Typically, a sample will require hundreds of runs to fully examine its parameter space. The most common scan is a rectangular scan of the 2D scanning mirrors, which moves the laser beamspot spatially over the surface of the sample, observing the photoresponse. These spatial scans are designed to be high resolution with variable speed, so that resolution in space can be traded-off for resolution in other variables when needed, while still spatially imaging. Depending on the measurement, the laser can be scanned in two dimensions, made to scan along a line in a single spatial dimension or held at a fixed position on the sample. Normally, when measuring begins on a new sample a set of low resolution runs are taken to determine the relevant parameters and the ranges that they vary over. Then a high resolution set of data cubes is taken to densely sample the full parameter space, commonly spatial scans as a function of two parameters, generating a large set of images that is usually the main result. Finally, if any unusual or interesting features are seen in that data set, some high resolution scans are taken to finely characterize those features, often continuing until the sample dies or degrades beyond usefulness.

\section{Hierarchical Analysis of Multi-Parameter Data Sets}

\begin{figure*}[t]
\centering
\includegraphics{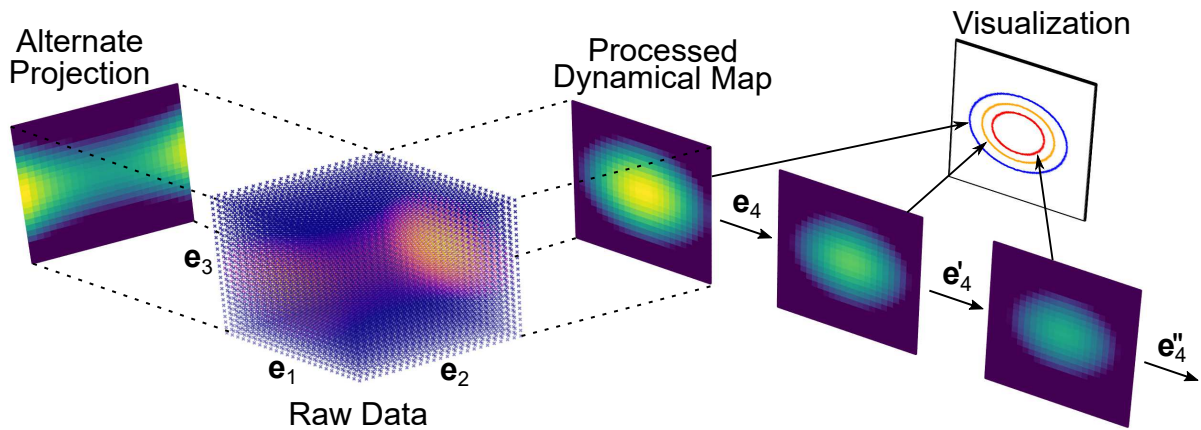}
\caption{A schematic of the analysis of a hypothetical four dimensional (hypercubic) data set, going from raw data to processed dynamical maps to a compact visualization.}
\label{fig:hypercube_process}
\end{figure*}

MPDPM generates large sets of images varying across several experimental variables, requiring sophisticated analysis to extract and visualize results. While the analysis of these sets will vary based on the specific sample, in this section we will give a general procedure to hierarchically exploit data geometry in order to condense a multivariate data set down to a manageable amount of processed data. Figure \ref{fig:hypercube_process} illustrates this process for a hypothetical four dimensional (hypercubic) data set. The raw data is a set of datacubes spanning three dimensions ($\mathbf{e}_{1}$, $\mathbf{e}_{2}$, $\mathbf{e}_{3}$), incremented along a fourth dimension $\mathbf{e}_{4}$. Each datacube is processed to map out a dynamical parameter that represents the behavior of the datacube along one axis (in this case $\mathbf{e}_{2}$). There are multiple possible projections and representations, although not all are useful the possibility space should be explored. Image analysis is used on the dynamical maps to identify key features that are then collected into a single visualization. In this hypothetical case ellipses enclosing the ``bright" photoresponse are visualized as contours. Ideally, this visualization will represent the evolution of some physically interesting quantity within the four-dimensional parameter space.

To illustrate this process in a vdW heterostructure device, Figure \ref{fig:analysis_diagram} presents data and analysis from a graphene on boron nitride on graphite (GBNGr) stacked heterostructure photocell, shown schematically in Figure \ref{fig:device}. The component materials were exfoliated from high quality bulk crystals onto Si/SiO$_2$ substrates. The heterostructure was assembled in a custom built transfer microscope using a well know dry transfer technique developed by Gomez et al.\cite{Gomez_Dry_Transfer_2014} Polydimethylsiloxane (PDMS)/polypropylene carbonate (PPC) stamps were used to pick up and controllably deposits the exfoliated flakes on top of each other. Titanium-gold (Ti/Au) electrical contacts were fabricated onto the device using electron beam lithography to provide electrical connection to the graphene (on the top) and graphite (on the bottom). When photoexcited, a Fermi-Dirac distribution of hot carriers rapidly forms in the graphene layer and the exponential tail of this hot distribution may extend into the valence band of the boron nitride, resulting in interlayer photocurrent between the graphene and graphite.\cite{Jarillo_Ma_Tuning_Thermalization_Pathways_Heterostructure_2016} To measure this interlayer photocurrent, the graphene contacts were set at a fixed voltage, and current was collected on the graphite and measured with a lock-in amplifier.

Using MPDPM on the GBNGr sample, we obtain our main data set: photocurrent data cubes composed of 25 spatial scans at varying laser power, repeated as a function of voltage (applied to the top graphene) in 2~mV increments from -20~mV to 30~mV, for a total of 625 spatial photocurrent images, sampling a four dimensional parameter space (two spatial dimensions, laser power and voltage). Following the general procedure, we will condense the data by fitting it to a phenomenological power law that describes the sample's behavior, identify a physically interesting nodal feature in the resulting non-linearity maps, and visualize the sample's behavior by tracking that node as a function of space. The GBNGr data is instructive because it has very distinct features in the non-linearity dynamical parameter, making analysis straightforward, but this approach can easily be adapted to other experiments and MPDPM data sets.

\subsection{Data Processing and Dynamical Fitting}

\begin{figure*}[t]
\centering
\includegraphics{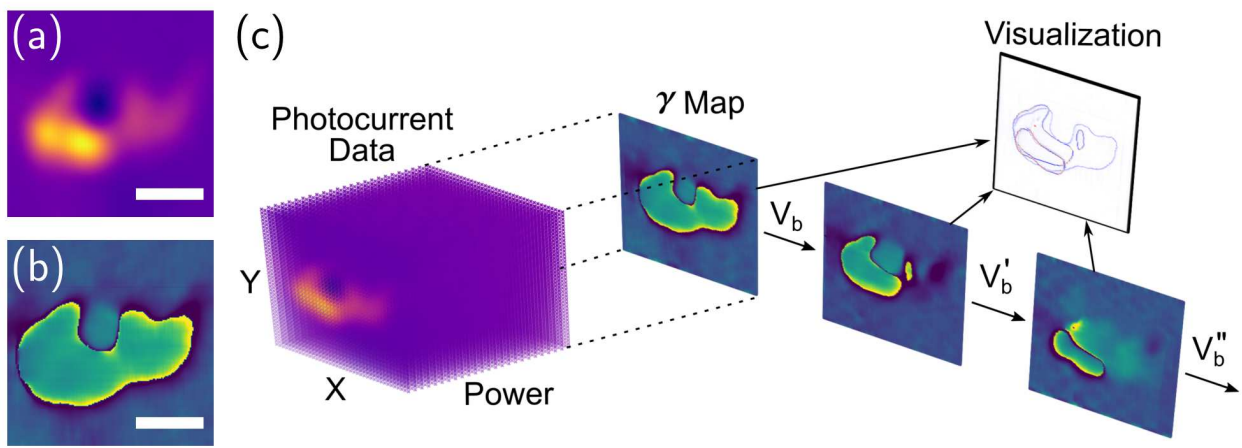}
\caption{(a) Shows an example photocurrent map from the GBNGr data set, and (b) shows an example $\gamma$ map that can be obtained by fitting data cubes of photocurrent images. Scale bars are 3 $\mu$m. (c) Schematic showing how the photocurrent data is condensed down into a set of $\gamma$ maps, which can then be further analyzed and collected into a visualization.}
\label{fig:analysis_diagram}
\end{figure*}

Data processing systematically prepares the raw MPDPM data and extracts dynamical variables that indicate changes in behavior. We use a set of custom python modules, together forming a ``toolbox" to handle data runs in a systematic manner. The lower right section of Figure \ref{fig:dataflow} shows the main functions of the toolbox. Given a run number, the code retrieves the relevant calibration data and returns the calibrated data along with all the experimental parameters. The next step is to combine the two dimensional images into a larger data set, such as constructing a three-dimensional data cube from a series of images. For spatial images, the image processing must account for the physical drift in the images, or other similar distortions, to spatially correlate the images.

The next step is to extract fitting parameters that can represent the dynamics occurring in the system. Once the data cube is spatially correlated, the data points are fit to a phenomenological law using a non-linear least squares fitting algorithm. The phenomenological law can be any function that parameterizes the data well. For photocurrent systems, we most commonly we use equations $I \propto P^\gamma$ and $I \propto e^{- \Delta t / \tau}$, for the photocurrent ($I$), versus laser power ($P$) or versus two pulse delay $\Delta t$, respectively. Phenomenological parameters, such as $\gamma$ and $\tau$, are extracted from these curve fits. These parameters should be dynamical quantities, so that they can represent changes in the underlying physics. For example, $\gamma$ is related to the non-linearity of the photoresponse, similarly $\tau$ is the characteristic timescale of a process. Changes to $\gamma$ or $\tau$ indicate a change to the character of the photoresponse, not simply a re-scaling of the data, making these parameters very useful proxies for the underlying physical phenomena.

The dynamical fitting parameters are used to condense the data. For the GBNGr sample the raw data consists of a set of photocurrent images as a function of power, one such image is shown as a colormap in Figure \ref{fig:analysis_diagram}(a). These images are correlated together, then the data at each point in space is curve fit along the power axis to the power law $I \propto P^{\gamma}$. This power law describes the data well in this case, and the parameter $\gamma$, acts as a index of the non-linearity, a useful dynamical quantity. The fitting gives a map of the fit parameter $\gamma$ as a function of space, such as that shown in Figure \ref{fig:analysis_diagram}(b). The processed $\gamma$ image condenses the dynamics of the whole three dimensional set of photocurrent images into a two dimensional map. The entire GBNGr data set is four dimensional, with data cubes taken as a function of laser power at various values of applied voltage. Figure \ref{fig:analysis_diagram}(c) shows how the data set is processed, all of the data cubes in the set are processed into $\gamma$ images, giving a three dimensional set of $\gamma$ images representing the sample non-linearity as a function of voltage. The resulting set of $\gamma$ maps can then be analyzed using image analysis to condense them into a single visualization.

\subsection{Image Analysis}

Once processed, image analysis is used to identify, and algorithmically extract, physically interesting features from the processed images. Identified features can be projected onto the spatial axes (or taken as a function of some other variable). This further reduces the dimensionality, usually giving a result that is visualizable as data mapped in space, or even as a function of a single variable, which human intuition is more suited to handle. The algorithm used to perform image analysis is the most application specific component of the process, as the ability to quantitatively pick a feature out of an image depends highly on what features are present. However, there are many well established image processing algorithms, and a researcher with a solid foundation in programming and signal processing should be able to find a solution without much trouble. In the author's experience, image filtration and basic optimization algorithms are usually all that is needed.

\begin{figure}[t]
\centering
\includegraphics{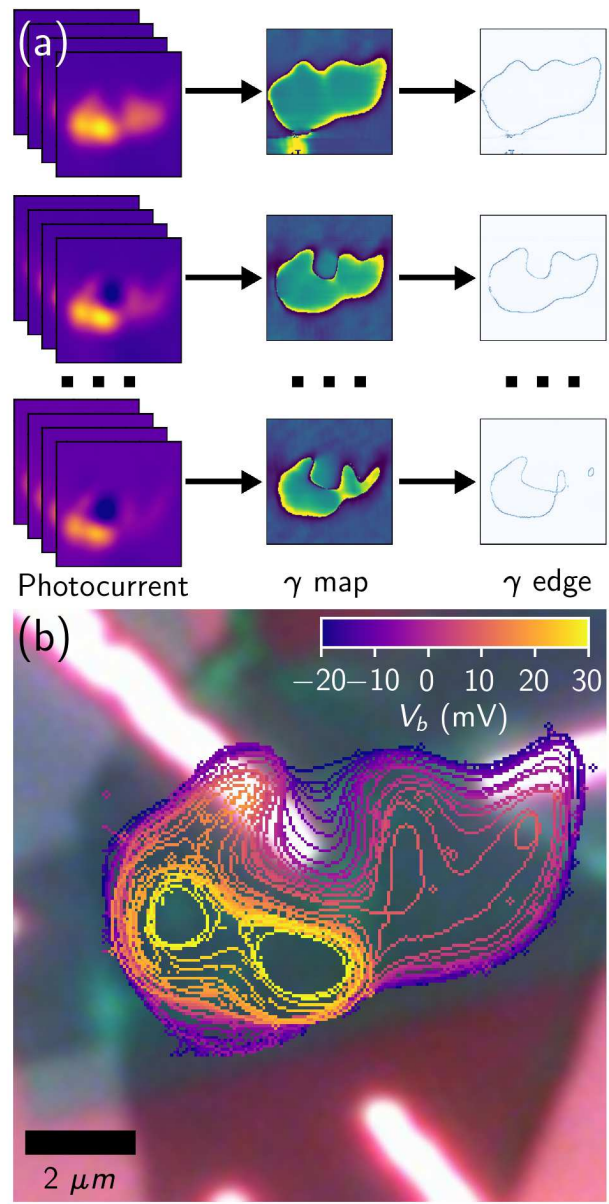}
\caption{(a) Schematic of the hierarchical analysis; processing data cubes into single dynamical $\gamma$ images, then picking out the key edge feature from the dynamical images. (b) The result of hierarchical analysis, overlaid on a high contrast optical image of the GBNGr sample. Physics can be determined from this visualization by interpreting the node as the zero point in the electrochemical potential of the sample.}
\label{fig:gamma_result}
\end{figure}

The GBNGr data provides a clear example of how to use image analysis to identify interesting features from image data. In the processed $\gamma$ maps there are distinct regions with different $\gamma$ values. The higher values of $\gamma$, (the green and yellow areas on Figure \ref{fig:analysis_diagram}(b)) are separated from the lower values of $\gamma$ and the background (blue and dark purple areas) by a sharp boundary. The boundary is a physically interesting feature, because it indicates a node in the photocurrent versus power, which evolves as a function of applied voltage due to the internal electronic properties of the sample. We use a Laplace filter, a common image processing filter used for edge detection, to identify this feature. This is performed on each $\gamma$ map at different values of applied voltage. The image analysis process is shown schematically in Figure \ref{fig:gamma_result}(a), the raw datacubes yield maps of $\gamma$ condensing the four dimensional data set down into a three dimensional data set. Then the edge feature is extracted from each $\gamma$ map, forming a highly condensed set of images showing only the feature of interest.

From the condensed data, which can be correlated to physical features of the sample, we can now develop an interpretation of the MPDPM data set. Figure \ref{fig:gamma_result}(b) shows the node versus applied voltage overlaid on an optical image of the GBNGr heterostructure sample. The edge is a node in the photocurrent, implying that charges excited at that location do not experience any force that would drive a current. This means that, on the node, the internal electrochemical potential of the sample is zero. Figure \ref{fig:gamma_result}(b) shows how the internal electrochemical potential of the sample is modified by an externally applied electric field. Of particular note, is the dipole-like feature in the top center of the nodal pattern which lies on top of an electrically floating metal contact. It has been predicted that a floating contact would modify the internal potential of graphene in a dipole pattern.\cite{Song_Shockley_Ramo_photocurrent}  

It would have been difficult to observe this data without using MPDPM. No single image, or dependence of a single parameter, contains a clear experimental signature of changing electrochemical potential. Only by sampling several experimental variables, observing the changing dynamics, and picking the right feature out of the complex photoresponse, could we identify this. In addition, there is no reason that the $\gamma$ node was the only interesting feature in the data set. In other studies, the authors have examined MPDPM sets with multiple different image analysis approaches, gleaning multiple physics results from a single MPDPM data set.

\subsection{Visualization}

Consistent and well considered visualization is important throughout the data analysis process. Developing a consistent way to visualize the data can prevent a researcher from becoming overwhelmed by the volume of data, and provide a platform for deeper forms of data analysis, but care must be taken as some visualizations can inhibit understanding. The use of colorscales in MPDPM is a good example of this, as colorscales are important when working with images but for some colorscales the non-uniformity of human color perception leads to perceived differences where there are none.\cite{Ware_Colormaps_1988, Moreland_Colormaps_Sci_Visualization} If this happens, a researcher combing through a large set of images they may waste time pursuing differences in contrast that appear to be significant, but aren't. A better way would be to utilize perceptually uniform colorscales, which are designed such that equal steps in data are perceived as equal color differences. In this work, we always use the matplotlib \textit{plasma} colorscale to represent photocurrent and the \textit{viridis} colorscale to represent $\gamma$ and other dynamical quantities.

More generally, visualizing data sets of more than two dimensions requires careful consideration, especially when it influences choices made in the analysis process. MPDPM data processing requires the choice of a fit function with a significant dynamical parameter, and MPDPM image analysis requires identifying an interesting feature to track. Intermediate visualizations of the data are needed to make these choices. To find the appropriate fit function we developed a set of python scripts that can consistently visualize cuts through the data sets looking for non-trivial functional dependence of the data. Once the fitting function is identified and dynamical parameter maps are calculated the (often three-dimensional) processed data must be visualized to identify key features for image analysis. While 3D renderings of the data, such as that shown in Figure \ref{fig:phase_space}(b), can be useful, they are heavily influenced by perspective, which can obscure details. In the author's experience it is best to examine three-dimensional data as a movie, a series of two-dimensional dynamical images with time representing the third axis. In this representation human perception is good at noticing changes, which lends itself well to identifying features that evolve as a function of time. Without consistent and well considered visualization it would be difficult to pick the right feature and results could be missed.

\section{Conclusions}

MPDPM is an efficient way to explore the complex behavior of 2D vdW Heterostructures and quantum metamaterials in general. MPDPM combines optical techniques that can excite complex and correlated behavior in 2D systems with an integrated data acquisition system that can make efficient tradeoffs between the resolution of parameters in order to sample a multivariate parameter space in a reasonable amount of time. Rather than sampling a single part of the parameter space, MPDPM takes a comprehensive approach, which means that complexity in the sample's photoresponse becomes complexity in the data. Therefore, advanced data analysis techniques are crucial to making MPDPM work. Fortunately, the high density and data geometry allow the data to be condensed and physically interesting features to be extracted and visualized. 

The number of different 2D materials and quantum metamaterials that can be fabricated is increasing rapidly, and so is the diversity of phenomena that they involve. As the field expands, it is important that researchers be able to comprehensively characterize their nanodevices, and as the complexity of those devices increases, the need for data intensive methods becomes greater. The general technique of MPDPM can be adapted to many other kinds of optical experiment beyond photocurrent, including, but not limited to, transmittance, photoluminescence, or photovoltage. The general idea of comprehensively and efficiently searching a parameter space is important for the discovery of new physics in these materials. Developing automated experimental systems reduces some of the burden and time limitations on researchers that would normally prohibit comprehensive characterization. In the future it may be possible to expand this further, developing search algorithms to explore sample parameter space and identify new phenomena with minimal human input.

\section{Acknowledgments}

We acknowledge Dennis Pleskot who fabricated the GBNGr device used as an example in this work, as well as the other members of the QMO lab at UCR for valuable discussions. This work was supported by the Air Force Office of Scientific Research Young Investigator Program (YIP) award \#FA9550-16-1-0216, as part of the SHINES center, an Energy Frontier Research Center funded by the U.S. Department of Energy, Office of Science, Basic Energy Sciences under award no. SC0012670, and through support from the National Science Foundation Division of Materials Research CAREER award no. 1651247. N.M.G. also acknowledges support through a Cottrell Scholar Award, and through the Canadian Institute for Advanced Research (CIFAR) Azrieli Global Scholar Award. In addition, support for T.B.A., as well as many valuable discussions with researchers of diverse data science backgrounds, was provided by the Fellowships and Internships in Extremely Large Data Sets (FIELDS) program, a NASA MUREP Institutional Research Opportunity (MIRO) program, grant number NNX15AP99A.

\bibliography{references_MPDPM}

\end{document}